# Maximum Spanning Tree Model on Personalized Web Based Collaborative Learning in Web 3.0

S.Padma [1] , Dr. Ananthi Seshasaayee [2]

[1] Research Scholar, Bharathiar University, Coimbatore
`padmanivasan@gmail.com`
[2] Head, Quaid-e-Millath Govt. College for women, Chennai
`ananthiseshu@gmail.com`

## ABSTRACT

Web 3.0 is an evolving extension of the current web environme bnt. Information in web 3.0 can be collaborated and communicated when queried. Web 3.0 architecture provides an excellent learning experience to the students. Web 3.0 is 3D, media centric and semantic. Web based learning has been on high in recent days. Web 3.0 has intelligent agents as tutors to collect and disseminate the answers to the queries by the students. Completely Interactive learner's query determine the customization of the intelligent tutor. This paper analyses the Web 3.0 learning environment attributes. A Maximum spanning tree model for the personalized web based collaborative learning is designed.

## 1. INTRODUCTION

Web 3.0 is an evolving extension of www, in which the information can be shared and interpreted by other software agent to find and integrate applications to different domains. Web 3.0 provides integrated real time application environment to the user. The applications are majorly involved in searching using semantic web, 3D web and are media centric. Web 3.0 supports pervasive components. Each component and its relations are represented below.

In web 3.0, web is transformed into database or Data Web wherein the data which are published in the web is reusable and can be queried. This enables a new level of data integration and application interoperability between platforms. It also makes the data openly accessible from anywhere and linkable as web pages do with hyperlinks. Data web phase is to make available structured data using RDF. The scope of both structured and unstructured content would be covered in the full semantic web stage. Attempts will be to make it widely available in RDF and OWL semantic formats.

The driving force for web 3.0 will be artificial intelligence. Web 3.0 will be intelligent systems or will depend on emergence of intelligence in a more organic fashion and how people will cope with it. It will make applications perform logical reasoning operations through using sets of rules expressing logical relationships between concepts and data on the web. With the realization of the semantic web and its concepts web 3.0 will move into Service Oriented Architecture.

The evolution of 3D technology is also being connected to web 3.0 as web 3.0 may be used on massive scale due to its characteristics. Web 3.0 is media centric where users can locate the searched media in similar graphics and sound of other media formats.

The pervasive nature of web 3.0 makes the users of web in wide range of area be reached not only in computers and cell phones but also through clothing, appliances, and automobiles.
Learning process in Web 3.0 is a highly sophisticated for the learners. The learners have access the unbelievable knowledge source . The tutors are intelligent agents who are customized for the learners.

## 2. REVIEW OF RELATED WORK

Claudio Baccigalupo and Enric Plaza discussed in the paper poolcasting : a social web radio architecture for Group Customization about Pool casting a social web radio architecture in which groups of listeners influence in real time the music played on each channel. Pool casting users contribute to the radio with songs they own, create radio channels and evaluate the proposed music, while an automatic intelligent technique schedules each channel with a group customized sequence of musically associated songs [13] .

M.T.Carrasco Benitez discussed in the paper Open architecture for multilingual social networking about an open architecture for all the multilingual aspects of social networking. This architecture should be comprehensive and address well-trodden fields such as localization, and more advanced multilingual techniques to facilitate the communication among users [14] .

Autona Gerber, Alta van der Merwe, and Andries Barnard discussed in the paper A functional Semantic web architecture about the CFL architecture which depicts a simplification of the original architecture versions proposed by Bernes-Lee as a result of the abstraction of required functionality of language layers. Gerber argues that an abstracted layered architecture for the semantic web with well defined functionalities will assist with the resolution of several of the current semantic web research debates such as the layering of language technologies [15] .

## 3. WEB BASED LEARNING - THE CURRENT SCENARIO

Web based learning is the electronically supported learning and teaching which are procedural in character and aim to effect the construction of knowledge with reference to individual experience, practice and knowledge of the learner. Information and communication systems, whether networked or not, serve as specific media to implement the learning process [1] .

The current web environment is on the transition state between web 2.0 and web 3.0. The Earlier Web based learning process was instructional. Using the internet technologies the students acquired knowledge. In Web 2.0 the learner is powered by the social softwares like blogs, wiki's ,

podcasts and virtual worlds. The value addition of social software has made the learner to be participatory in the learning process. Social networking software's are participatory knowledge acquisition phenomenon which shares the knowledge of the society.

A typical environment of learning in Web 2.0 is read and write .The primary participants are learner and a tutor .The learner can be an active participant in the social networking sites and acquire knowledge. When the learner opts for undergoing classes in a particular topic, he learns from the static knowledge of the tutor which are programs. It includes static contents and the learner can ask questions and answers if supported by the programs will be displayed. The tutors look and appearance is 2D.

The main disadvantage of Personalized learning in Web 2.0 is although the environment is participatory, it is static and predefined. It does not includes the collective knowledge of current advancements in the topic. The interface is simple to use. The Learner can have only limited queries. The dissemination of knowledge by the tutor is determined by the developer.

## 4. TECHNICAL BACKGROUND OF WEB 3.0

Web 3.0 architecture makes the learner sophisticated. Resource Definition Frame work (RDF), SPARQL, WebDAV, site Specific API's ,FOAF and SSL are the components of Web 3.0 Resource Definition Frame work is a meta data data model . The resources are transformed into statements of the form subject-predicate-object triple expressions. The subject denotes the resource, predicate denotes traits or aspects of the resource and express a relationship between the subject and the object. For example to represent the notion "The sky has the color blue" in RDF is as the triple: a subject denoting "the sky", a predicate denoting "has the color", and an object denoting "blue" . A collection of RDF statements intrinsically represents a labeled, directed multi-graph. As such, an RDF-based data model is more naturally suited to certain kinds of knowledge representation [2] . In web 3.0 Personalized learning is accomplished by associating the queries by learners and the knowledge of the tutor together with the available web resources in RDF format so that the query can be associated with the knowledge.SPARQL protocol for RDF query language process query which consist of triple patterns, conjunctions , disjunctions and optional patterns. SPARQL facilitates personalized web based learning in web 3.0 by accepting patterns, multiple queries and complicated queries by the learner [3] . It also empowers the intelligent agent tutor to collaborate multiple web resources into meaningful and understandable text and visuals of the learner.

Web-based Distributed Authoring and Versioning (WebDAV) is a set of methods based on the Hypertext Transfer Protocol (HTTP) for collaboration between users in editing and managing documents and files stored on World Wide Web servers. The most important features of the WebDAV protocol include:locking (overwrite prevention ),properties (creation, removal, and querying of information about author, modified date),name space management ( ability to copy and move Web pages within a server's namespace )and collections (creation, removal, and listing of resources) [4] .

In Personalized collaborative environment webDAV assists the intelligent agent tutor to interact with the web servers to collect and produce cognizant knowledge to to the learner.
Site specific API's interacts with the RDF and produces required output of the learner. Site specific API's provides the functionality to the learner by making the query understandable to the tutor in precise form. It supports the intelligent agent tutor to interact with the web data resources in world wide web servers to collaborate the data [5] .

Site Specific API's are useful while modifying the existing personalized learning web sites of web 3.0 so that the functionality can be reused and remixed again. FOAF+SSL is a decentralized secure authentication protocol using the FOAF profile information and SSL security layer. It makes ease the role of the intelligent agent tutor in web 3.0 by maintaining FOAF cloud of the persons , activities their relations to other people and objects [5] . The intelligent agent tutor of web 3.0 need not maintain a centralized machine readable ontology . The security issues are maintained by SSL.

The aggregation of these components is utilized in the personalized collaborative learning of web 3.0. Technology plays a vital role in all aspects of personalized learning. The technologies which are utilized in personalized collaborative learning includes Artificial intelligent, Automated reasoning, Cognitive architecture, Composite applications, Distributed computing, Knowledge representation, Ontology, cloud computing, Gridcomputing, Scalable vector graphics, Semantic web, Semantic WiKi and Software agents.

## 5 . PERSONALIZED WEB BASED LEARNING IN WEB3.0

In web 3.0 the two major components are the learner and the 3D tutor. The learner is the human who intends to acquire knowledge about a specific subject. The tutor is an intelligent agent which delivers the collective knowledge to the learner.

The learner initially starts to take up the tutorial. The learner specify the tutor's look . He can specify the gender of the tutor. The look or avatar of tutor can also be specified by the learner. During the learning session, the learner must be completely interactive. His interactive queries , interpretations and examples defines the intelligence of the tutor. The learner can request more examples , working 3D models , clarifications, justifications, applications at any point of time. The output can be audio, video, 3D, or text as opted by the learner. It can be a combination of the output formats also depending on the preferences of the learner.

The learner can demand aggregations and mash up of websites based on the topic. He can have choice of collaboration of best web resources and technologies available till date.

The tutor's look is specified by the user. The tutor intuitively asses the intelligence of the learner. Based on the intelligence and personal preferences of learner the tutor must deliver the knowledge. The tutor must collaborate knowledge from various web resources , filter the irrelevant knowledge and share it. The tutor must give outputs according to the personal preferences of the learner and also prefer the alternative output form if one type is not sufficient. It must understand the language of the learner and interact accordingly. The agent must aggregate the web resources required by the learner which is updated till date. The technologies and tools used must be the suitable by analyzing all the till date available technologies and tools

## 6. ATTRIBUTE ANALYSIS OF PERSONALIZED WEB BASED LEARNING IN WEB 3.0

Attributes in web 3.0 are the characteristics which determine the behaviour of the applications in web 3.0 architecture. Attribute analysis determines the key factors in developing web 3.0 applications. The attributes considered are 3D, Mash ups, Preferences regarding e-tutor and Speech recognition.

## 7. DATA COLLECTION

Data collection is a process of preparing and collecting data [10] . The method of Data collection adopted is Questionnaire. A Questionnaire is designed with a closed type question. Responses of the questionnaire is got from 100 Post Graduate Compute stream Students at a private university in Chennai.

The questionnaire sample is given below.

Questionnaire :

| s.no | Question | Response type |
| --- | --- | --- |
| 1. | Do you require a 3D tutor ? | Yes/no |
| 2. | Can the Demos for your learning be in 3D? | Yes/no |
| 3. | Do you wish to ask a knowledge based random query to the etutor ? | Yes/no |
| 4. | Will you prefer a collective text output ? | Yes/no |
| 5. | Do you want a collective knowledge of the web sites of your choice ? | Yes/no |
| 6. | Do you want the tutor to teach according to your knowledge level ? | Yes/no |
| 7. | Will you like to opt the gender of your tutor ? | Yes/no |
| 8. | Should your tutor know about your preferences ? | Yes/no |
| 9. | Should the tutor have a collective knowledge of the best sites available ? | Yes/no |
| 10. | Do you wish your speech to be recognized by your tutor ? | Yes/no |

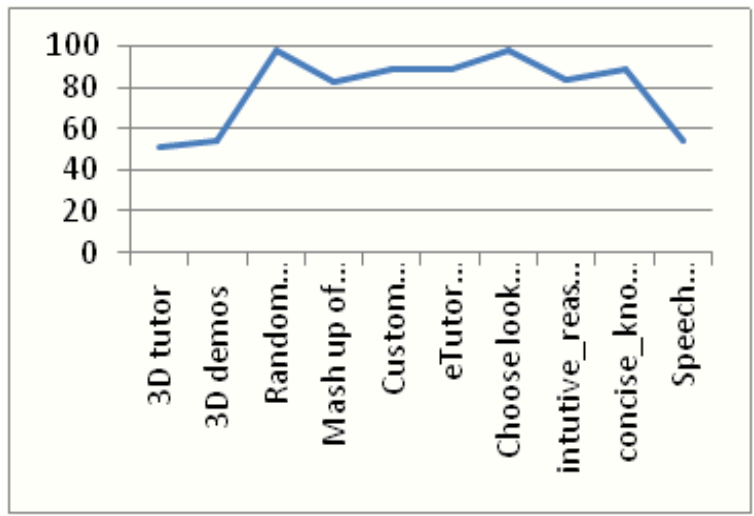

**Figure 1 : Sample input**

**Figure 2 : Comparison of requirements**

**The Top requirements are** Random question, Custom Mash up, Etutor teaches according to the level Choose look of etutor, concise_knowledge.

## 8. FINDING THE FREQUENT ITEM SETS IN THE ATTRIBUTES CONSIDERED USING APRIORI ALGORITHM

One of the most popular data mining approaches is to find frequent itemsets from a transaction dataset and derive association rules. Finding frequent itemsets (itemsets with frequency larger than or equal to a user specified minimum support) is not trivial because of its combinatorial explosion. Once frequent itemsets are obtained, it is straightforward to generate association rules with confidence larger than or equal to a user specified minimum confidence.

| A-Priori parameters | |
|---|---|
| Support min | 0.50 |
| Confidence min | 0.75 |
| Max rule length | 2 |
| Lift filtering | 1.00 |

**Figure 3. Apriori parameters**

Apriori is a seminal algorithm for finding frequent itemsets using candidate generation [ 12 ]. It is characterized as a level-wise complete search algorithm using anti-monotonicity of itemsets, "if an itemset is not frequent, any of its superset is never frequent". By convention, Apriori assumes that items within a transaction or itemset are sorted in lexicographic order. Let the set of frequent itemsets of size $k$ be $F_k$ and their candidates be $C_k$ . Apriori first scans the database and searches for frequent itemsets of size 1 by accumulating the count for each item and collecting those that satisfy the minimum support requirement. It then iterates on the following three steps and extracts all the frequent itemsets [11] .

Apriori algorithm was adopted to the data set. Let R1 represent Random question, R2 represent Custom Mash up, R3 represent Etutor teaches according to the level ,R4 represent Choose look of etutor, R5 represent concise_knowledge.

## 9. REQUIREMENT MATRIX :

A requirement matrix is constructed based on the analysis done using apriori algorithm.

|    | R1 | R2 | R3 | R4 | R5 |
|----|----|----|----|----|----|
| R1 | 0  | 0  | 0  | 1  | 0  |
| R2 | 0  | 0  | 1  | 0  | 1  |
| R3 | 0  | 1  | 0  | 0  | 1  |
| R4 | 1  | 0  | 0  | 0  | 0  |
| R5 | 0  | 1  | 1  | 0  | 0  |

**Figure 4. Requirement matrix**

From the matrix representation it is evident that,
{ R1 => R4}
{R2 => R3 ,R5}

International Journal of Computer Science, Engineering and Information Technology (IJCSEIT), Vol.1, No.5, December 2011

{R3=>R2,R5}
{R4=>R1}
{R5=>R2,R3}
R2,R3,R5 are strongly inter related requirements
R1,R4 are strongly inter related requirements

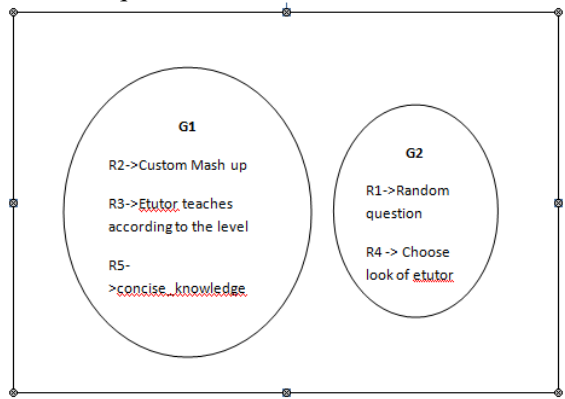

**Figure 5. Classified Groups**

## 10. CORRELATION AMONG THE ATTRIBUTES AND GROUPS

The correlation coefficient for all pairs among the Groups are calculated using the following formula.[7]

Correlation(r) =[ NΣXY - (ΣX)(ΣY) / Sqrt([NΣX2 - (ΣX)2][NΣY2 - (ΣY)2])]

where
    N = Number of values or elements
    X = First Score
    Y = Second Score
    ΣXY = Sum of the product of first and Second Scores
    ΣX = Sum of First Scores
    ΣY = Sum of Second Scores
    ΣX2 = Sum of square First Scores
    ΣY2 = Sum of square Second Scores

sample correlation coefficient of G1

| *S.No* | *Source* | *Destination* | *Correlation Coefficient* |
|---|---|---|---|
| 1. | R2 | R3 | .148 |
| 2. | R2 | R5 | .198 |

Table 1 : Correlation Coefficient

## 11. MAXIMUM SPANNING TREE ALGORITHM

A spanning tree of an undirected graph of n nodes is a set of n − 1 edges that connects all nodes. This note develops two algorithms for finding the minimum spanning tree. Properties of spanning trees In a spanning tree:

• There is no cycle: a cycle needs n edges.
• There is exactly one path between any two nodes: there is at least onepath between any two nodes because all nodes are connected. Further,there is not more than one path between a pair of nodes because thenthere would be a cycle that includs both nodes.
• Adding a non-tree edge creates a cycle: Suppose a non-tree edge (x, y) is added to a spanning tree. Now there are two distinct paths between(x, y), the added edge and the path in the tree. Hence there is a cycle.
• Removing an edge from a cycle as above creates a spanning tree: after removal of the edge there are (n − 1) edges. All nodes of the graph are connected: suppose edge (x, y) is removed that belonged to the original graph. The nodes x, y are still connected because x, y were on a cycle. For other node pairs, in the path in the original graph replace the edge (x, y) by the path between x, y.

**Kruskal's Algorithm**

Let $G = (V, E)$ be the given graph, with $|V| = n$

{

Start with a graph $T = (V, \phi)$ consisting of only the vertices of $G$ and no edges;

/* This can be viewed as $n$ connected components, each vertex being one connected component */

Arrange E in the order of increasing costs;

**for** ($i = 1, i<=n - 1, i + +$)

{

Select the next biggest cost edge;

**if** (the edge connects two different connected components)

add the edge to $T$;

}

}

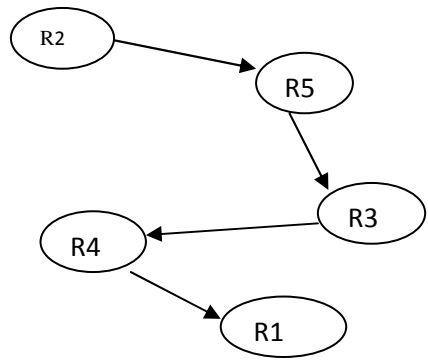

**Figure 6. Maximum spanning tree model**

The Maximum cost obtained is 2.238

## 12 CONCLUSION

From the analysis ,it is concluded that Custom Mash up, Etutor teaching according to the level of the learner ,and concise_knowledge of the tutor are the interrelated requirements by the students for designing a tutorial web3.0 site. Learner asking Random question and learner choosing the look of etutor are the interrelated requirements by the students for designing a tutorial web 3.0 site. If any one of the interrelated requirement are required by the specific clients , then the developer can include all the inter related requirements in the a tutorial web 3.0 product for better customer satisfaction. Also Custom Mash up , concise_knowledge , concise_knowledge , Choose look of etutor, Random question is the feature order which should be given preference  in the design and development of web 3.0 tutorial site.

**Authors**

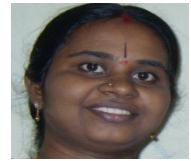

**S.Padma**, is a research scholar in Bharathiar university ,Coimbatore. She has published 4 international journals .Currently she is working as Assistant professor in School of Computing sciences, Vels university, Chennai , India. Her area of interest is web mining.

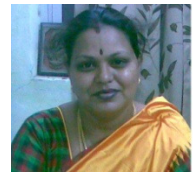

**Dr. Ananthi Seshasaayee**  received her Ph.D in Computer Science from Madras University. At   present she is working as  Associate professor and Head, Department of computer science, Quaid-e-Millath Government College, for Women, chennai. She has published 19 international journals. Her area of interest involve the fields of Computer Applications and Educational technology.